\documentclass[aps,prl,twocolumn,groupedaddress,superscriptaddress,showpacs]{revtex4-1}

\usepackage{epsfig,amssymb,amsmath,graphicx,subfigure,hyperref}
\usepackage{tabularx}
\usepackage{float}
\usepackage{setspace}
\usepackage{color}
\usepackage{times}
\usepackage{verbatim}

\usepackage{array} 
\newcommand{\PreserveBackslash}[1]{\let\temp=\\#1\let\\=\temp} \newcolumntype{C}[1]{>{\PreserveBackslash\centering}p{#1}} \newcolumntype{R}[1]{>{\PreserveBackslash\raggedleft}p{#1}} \newcolumntype{L}[1]{>{\PreserveBackslash\raggedright}p{#1}} 
\usepackage{textcomp}

\begin{document}

\title{Spontaneous tilt of single-clamped thermal elastic sheets}

\author{Zhitao Chen}
\affiliation{Department of Physics, University of California Santa Barbara, Santa Barbara, California 93106, USA}
\author{Duanduan Wan}
\affiliation{School of Physics and Technology, Wuhan University, Wuhan 430072, China}
\author{Mark J.~Bowick}
\email{E-mail: bowick@kitp.ucsb.edu}
\affiliation{Kavli Institute for Theoretical Physics, University of California Santa Barbara, Santa Barbara, California 93106, USA}

\begin{abstract}
Very thin elastic sheets, even at zero temperature, exhibit nonlinear elastic response by virtue of their dominant bending modes. Their behavior is even richer at finite temperature. Here we use molecular dynamics (MD) to study the vibrations of a thermally fluctuating two-dimensional elastic sheet with one end clamped at its zero-temperature length. We uncover a tilt phase in which the sheet fluctuates about a mean plane inclined with respect to the horizontal, thus breaking reflection symmetry. We determine the phase behavior as a function of the aspect ratio of the sheet and the temperature.  We show that tilt may be viewed as a type of transverse buckling instability induced by clamping coupled to thermal fluctuations and develop an analytic model that predicts the tilted and untilted regions of the phase diagram. Qualitative agreement is found with the MD simulations. Unusual response driven by control of purely geometric quantities like the aspect ratio, as opposed to external fields,  offers a very rich playground for two-dimensional mechanical metamaterials.
\end{abstract}

\maketitle

Elastic sheets and cantilever ribbons have long been studied in classical plate theory \cite{Blevins1984, Leissa1969}. The energetic cost of bending, escape into the third dimension through height fluctuations, relative to elastic stretching is controlled by the dimensionless F\"oppl-von K\'arm\'an number vK $\sim \rm{A/t^2}$, where $\rm{A}$ is the area of the sheet and t is the thickness.  In the very thin limit, such as atomically-thin graphene, bending dominates and vK may be tuned by varying purely geometric scales, rather than external fields. The mechanical behavior of thin sheets is even richer when they are thermalized \cite{NelsonPiranWeinberg, Bowick&Travesset}. Instead of always crumpling, like linear polymer chains, thermal excitations lead to a low-temperature wrinkled flat phase, even for arbitrarily large sheets. In the wrinkled phase the bending rigidity and elastic moduli become scale-dependent rather than simple material parameters (see e.g. Ref.~\cite{Nelson&Peliti, Kantor1987, Aronovitz&Lubensky,guitter,LeDoussal&Radzihovsky, Kosmrlj&Nelson2016, Ahmadpoor2017, LeDoussal&Radzihovsky2018, Sajadi2018, Morshedifard2021}). In particular the bending rigidity, $\kappa$, is strongly scale-dependent, with an enhancement over the zero-temperature value by a factor  $(L/l_{\rm{th}})^\eta$, where $L$ is the smallest 2D spatial extent of the sheet, say the length $L$, $l_{th}$ is the length scale above which the effect of thermal fluctuations becomes significant, and the critical exponent $\eta$ is approximately 0.8 (e.g. Ref.~\cite{LeDoussal&Radzihovsky, Kownacki2009, Gazit2009, Hasselmann&Braghin, Troster2013,Troster2015}). This almost linear enhancement allows for further geometric tuning of the thermalized mechanical response, especially since the thermal length scale is of order nanometers or less for strong covalently bonded materials such as graphene \cite{Blees2015, Nicholl2015}. The combination of thermal fluctuations and geometric control provides a rich toolbox for generating unusual behavior. 
Here we show that a thermalized elastic ribbon of length $L_0$, clamped along only one edge of width $W_0$, like a miniature diving board and the standard setup for a cantilever, exhibits a transition in which it spontaneously tilts, that it oscillates about a mean tilted plane with respect to the horizontal, for a temperature-dependent range of aspect ratios $\alpha = (W_0/L_0)$.  Since the tilt plane is equally likely to be above or below the horizontal plane, we have in fact a 2-state oscillator, as in the case of thermalized Euler-buckling \cite{Hanakata}. 
We establish the tilt transition via Molecular Dynamics simulations and provide a theory
by reformulating it as a buckling instability resulting from clamping-induced strains with respect to the natural finite-temperature equilibrium state.

\begin{figure}
\centering
\includegraphics[width=2.8 in]{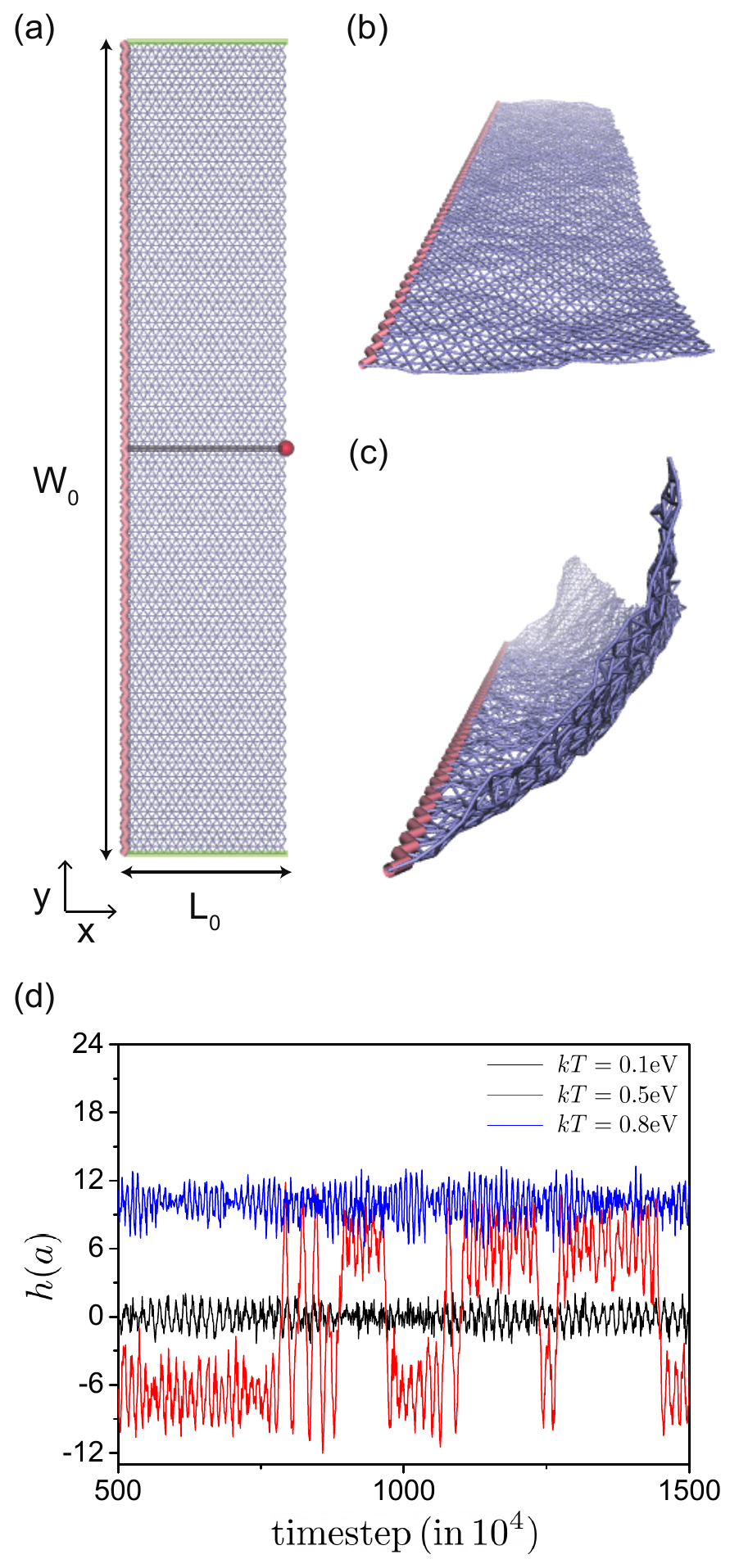}
\caption{(Color online) (a) A triangulated membrane with zero-temperature length $L_{0}=20a$ and aspect ratio $\alpha=W_{0}/L_{0} \approx 5$ clamped on the back edge (colored pink). The middle vertex on the front edge is marked with a large red dot. We label the left and right edges in green, and the centerline in grey. The snapshot was generated using the Visual Molecular Dynamics (VMD) package \cite{Humphrey1996} and rendered using the Tachyon ray tracer \cite{Stone1998}. (b) Snapshot of the horizontal phase. (c) Snapshot of the tilt phase. (d) Height of the red vertex as a function of time for $10^6$ time steps after equilibrating for $5 \times 10^6$ time steps.}
\label{fig:setup}
\end{figure}

We model an elastic sheet as a discrete triangular lattice of vertices and bonds, with the elastic energy a sum of stretching  and bending energies
\begin{equation}
E=\frac{\varepsilon}{2}\sum_{\left< ij \right>}\left(\left|\mathbf{r}_{i}-\mathbf{r}_{j} \right| -a \right)^{2}+\frac{\tilde{\kappa}}{2}\sum_{\left< I J \right>}\left(\mathbf{\hat{n}}_{I}-\mathbf{\hat{n}}_{J}\right)^{2}
\label{free_energy}
\end{equation}
where $\varepsilon$ is the discrete spring constant, $a$ is the equilibrium spring length and $\tilde{\kappa}$ is the discrete bending modulus. The $\left<ij\right>$ sum is over  pairs of nearest-neighbor vertices, with positions $\mathbf{r}_{i}$ in 3D Euclidean space, while the $\left<IJ\right>$ sum is over all pairs of triangular plaquettes, with unit normals $\mathbf{\hat{n}}_{I}$, that share a common edge. The continuum limit of Eq.~(\ref{free_energy}) leads to a Young's modulus $Y= 2\varepsilon/\sqrt{3}$, a bending rigidity $\kappa = \sqrt{3}\tilde{\kappa}/2$ and a Poisson ratio $\nu = 1/3$ \cite{Seung1988,Lidmar2003,Schmidt2012}. For graphene the discrete triangular lattice may be viewed as the dual of its actual honeycomb lattice \cite{Wan2017prb} with edge length $a = \sqrt{3}a_0$, where $a_0 = 1.42 \mbox{\AA}$ is the carbon-carbon bond length. Graphene's microscopic material parameters are $\kappa =1.2 \, \mbox{eV}$ \cite{Nicklow1972,Fasolino2008} and $Y=20 \, \mbox{eV}/\mbox{\AA}^{2}$ \cite{Lee2008,Zhao2009}.  Fig. \ref{fig:setup}(a) displays the zero-temperature flat configuration of a sheet in the $x-y$ plane, with $L_{0}=20a\approx 50\mbox{\AA}$ and aspect ratio $\alpha=W_{0}/L_{0} \approx 5$, where the subscript $0$ labels zero-temperature quantities. We clamp the edge vertices along one zigzag boundary indicated by the pink line in Fig.~\ref{fig:setup}(a) and tag the middle vertex on the free end (shown in red). We find consistent results from MD simulations using two different software packages: HOOMD-blue \cite{hoomd, Glaser2015} and LAMMPS \cite{Plimpton1995}. After giving the free vertices a small random out-of-plane displacement, we update their positions in the constant temperature (NVT) ensemble. With the mass, length and energy units chosen in simulations and the chosen integration time step, every simulation timestep $\tau$ corresponds to a real time $\tau \approx 0.6 \, \mbox{fs}$ (see Supplementary Material for details \cite{Supp}). Every simulation run consists of $10^7$ time steps in total, with the first $5\times 10^6$ time steps ensuring equilibration.

Since tilt occurs primarily for aspect ratios significantly above one we will use the terminology flap, rather than ribbon, for our elastic sheet. A flap exhibits two phases depending on the aspect ratio and the temperature: a horizontal phase where the flap vibrates about the horizontal $z=0$ plane, and a tilt phase where it vibrates about a tilted plane. We show snapshots of the two phases in Figs.~\ref{fig:setup}(b) and (c). It is revealing to  plot the height h (z coordinate) of the middle vertex of the free long edge (the red vertex in Fig.~\ref{fig:setup}(a)) for $10^6$ timesteps after equilibrating -- see Fig.~\ref{fig:setup}(d). At low temperature ($kT=0.1 \mbox{eV}$), the red vertex vibrates about $z=0$ (black line). At a higher  temperature ($kT=0.8 \mbox{eV}$), however, the vertex vibrates about $z \approx 10$ -- the upper trace (blue line). At an intermediate temperature ($kT=0.5 \mbox{eV}$), the vertex vibrates about two symmetric positions $z \approx \pm 7$ with occasional inversions (red line). 

\begin{figure}[H]
\centering
\includegraphics[width=3.2 in]{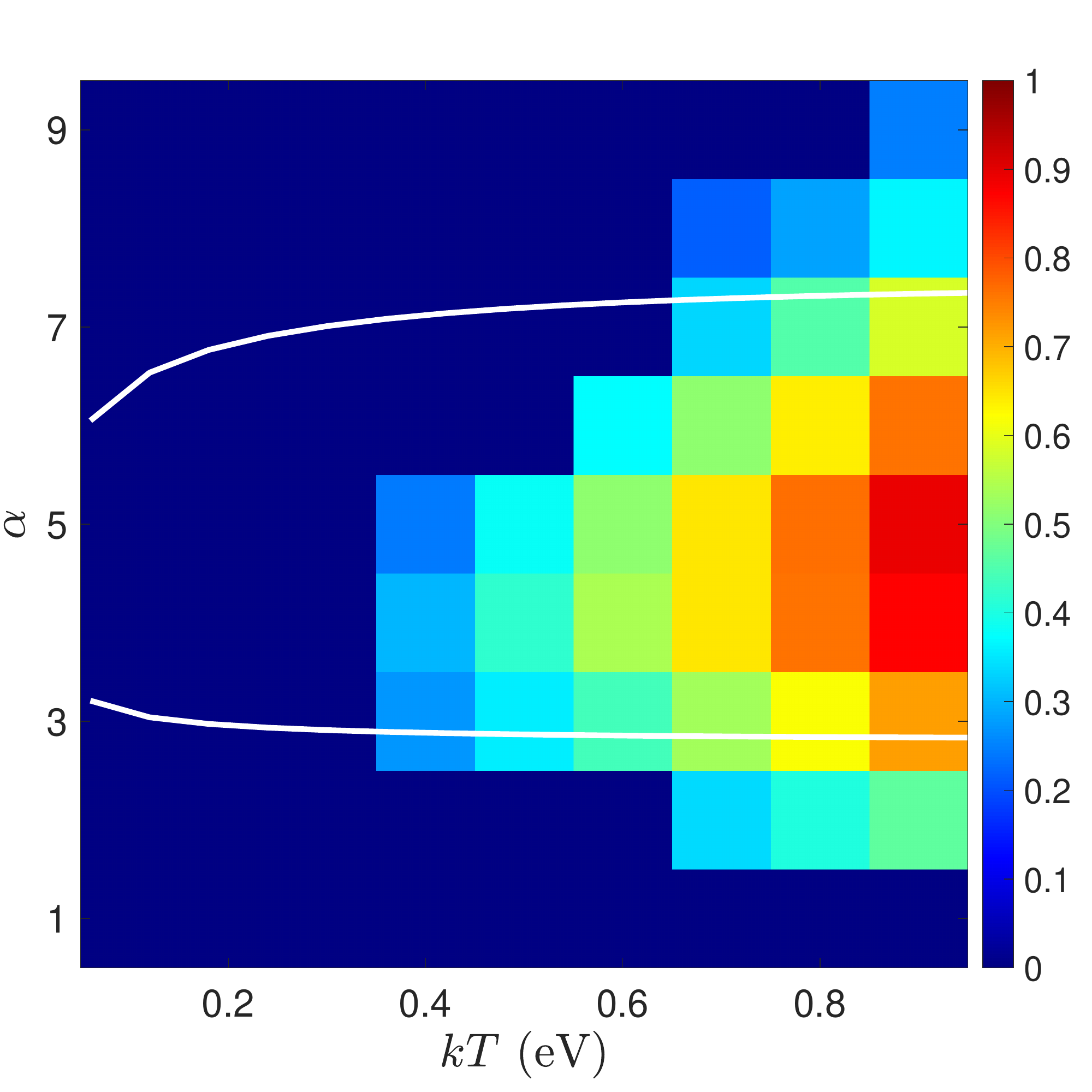}
\caption{(Color online) The value of order parameter $\phi$ as a function of temperature and aspect ratio $\alpha$. The diagram is obtained by analyzing the second $5\times 10^6$ time steps and averaging over five independent runs. White lines indicate estimated phase boundary by solving $\Delta_m=\Delta_c$, which are described in Eq.~(\ref{eq:delta}) and Eq.~(\ref{eq:delta_c}), respectively.}
\label{fig:order}
\end{figure}

\begin{figure}
\centering
\includegraphics[width=3.2 in]{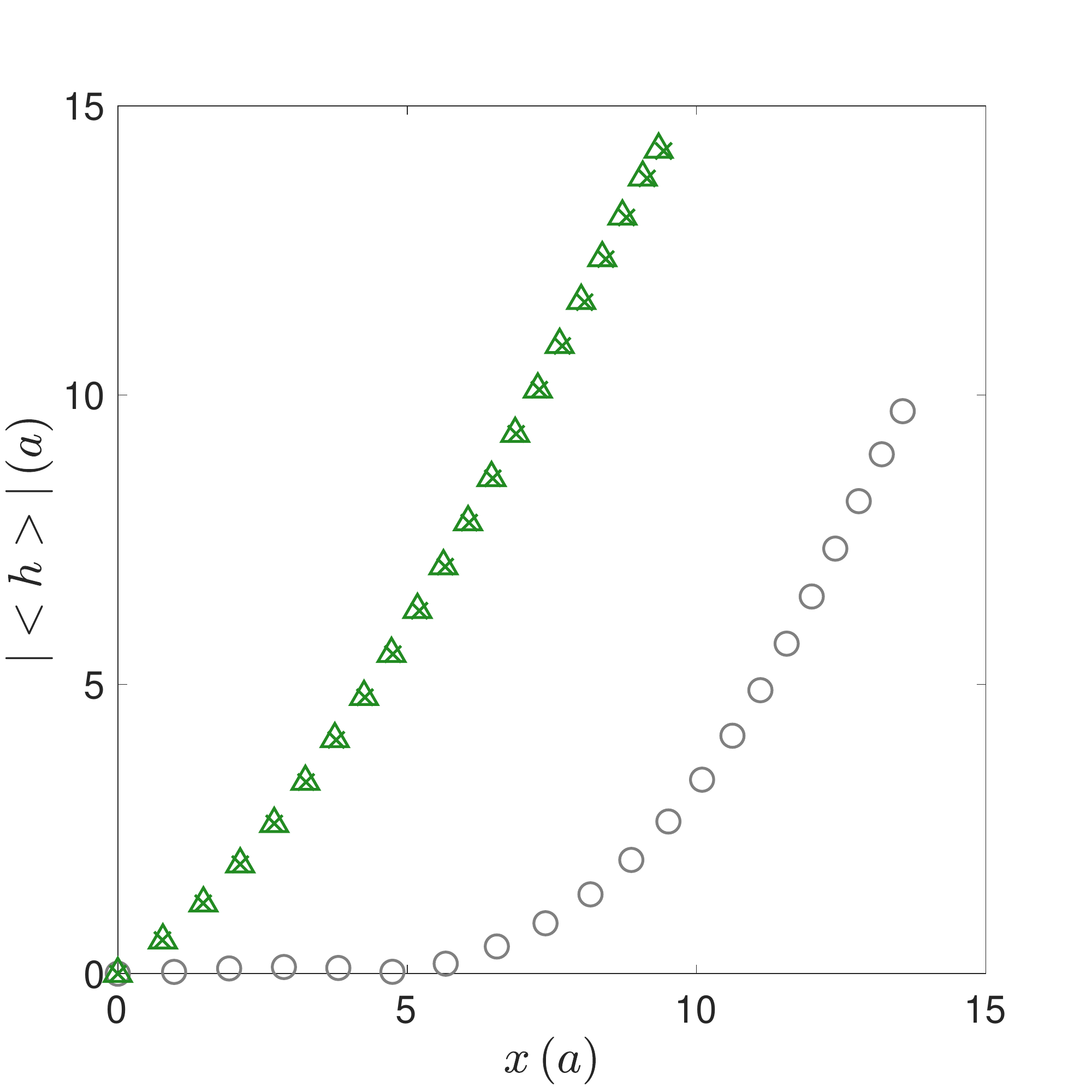}
\caption{(Color online) Profile of the two short free edges (marked in green in Fig.~\ref{fig:setup}(a)) and the parallel middle line (marked in grey in Fig.~\ref{fig:setup}(a)) of a tilted membrane with $kT=0.8 \mbox{eV}$ and $\alpha=5$. The green triangles correspond to the top and bottom edges, and the grey circles   correspond to the midline.}
\label{fig:profile}
\end{figure}

To distinguish the tilted phase from the horizontal phase, we measure the probability that $|h(t)|$ is $2.5$ standard deviations from $z=0$. We use a threshold probability of $1/4$, below which the flap is in the horizontal phase and above which it is in the tilted phase. To quantify tilt, we introduce an order parameter $\phi\equiv <|z/x|>$ within the tilted phase. In the horizontal phase $\phi$ is defined to be zero. We plot $\phi$ as a function of aspect ratio $\alpha$ and temperature $kT$ in Fig.~\ref{fig:order}, where we have averaged over five independent runs. At sufficiently high temperature and in an range of moderate aspect ratios, the flap is clearly tilted; at low temperature or outside the above window of aspect ratios the flap is horizontal. A close look at a typical tilt configuration shows that the flap is not uniformly tilted along the width direction. We plot the profile of the two short free edges (marked in green in Fig.~\ref{fig:setup}(a)) and the parallel middle line (marked in grey in Fig.~\ref{fig:setup}(a)) in the tilted phase in Fig.~\ref{fig:profile}. It can be seen that the two free edges tilt up straight while the middle line has a curved buckled profile: close to the clamped boundary, it does not deviate much from $z=0$; far away from the clamped boundary it tilts up straight.

\begin{figure}
\centering
\includegraphics[width=8.6cm]{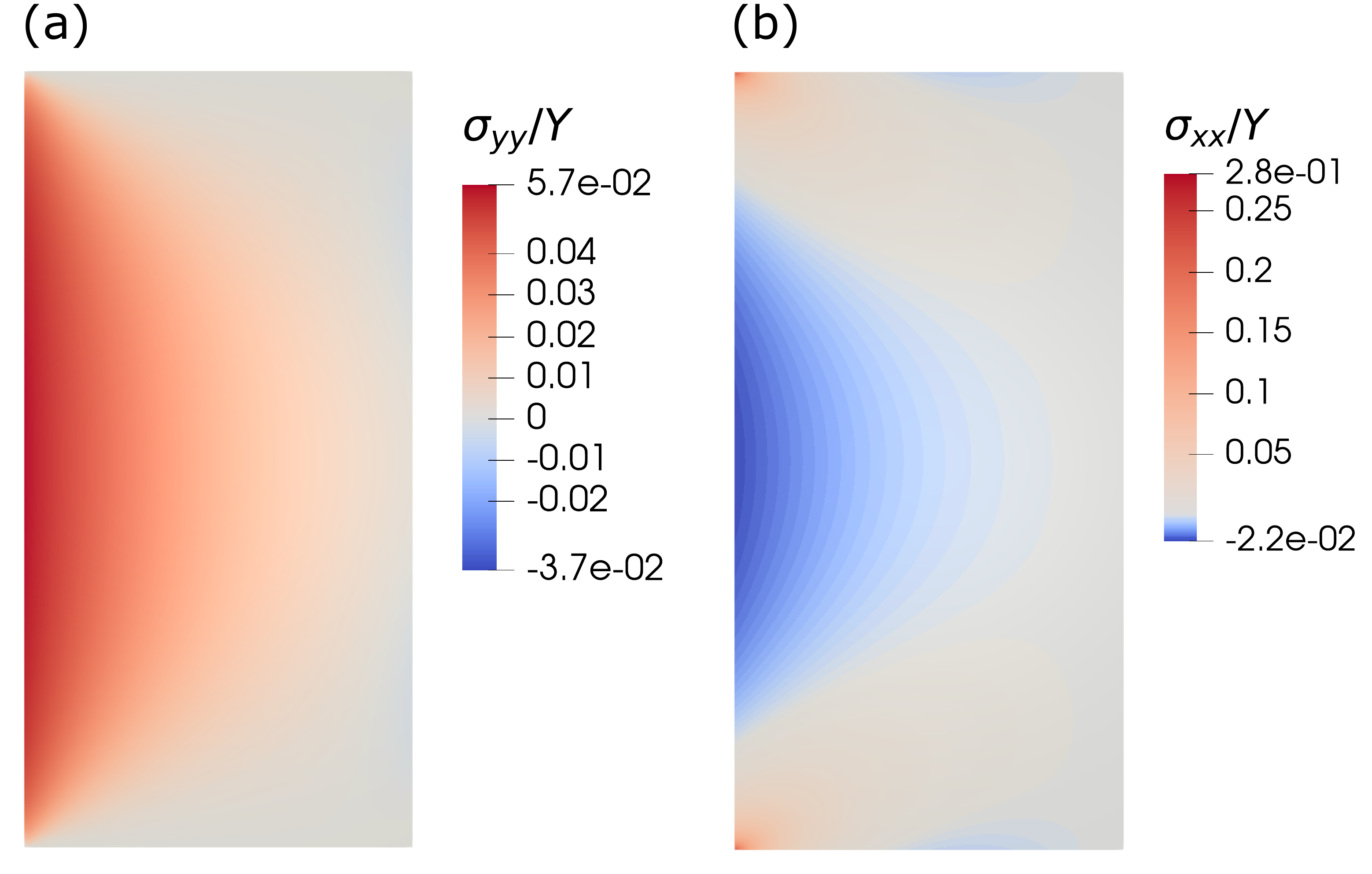}
\caption{(Color online) In-plane stress fields $\sigma_{yy}/Y$ and $\sigma_{xx}/Y$ of a rectangular sheet with $\alpha=2$ and is stretched by $\epsilon=0.05$ on the clamped edge (left edge). The finite element simulation is carried out using the FEniCS package \cite{fenics}.} 
\label{fig:stress}
\end{figure}

The tilt phase may be understood as a result of a buckling instability: a thermalized membrane has a projected area smaller than its zero-temperature area (e.g. Ref.~\cite{Kosmrlj&Nelson2016, Wan2017prb}). The natural reference state for defining stresses and strains is the thermalized membrane. Thus clamping one end at its zero-temperature width $W_{0}$ exerts a stretching force along the clamped boundary. Even at $T=0$ clamping has a measurable effect. Fig.~\ref{fig:stress} shows a rectangular sheet with $\alpha=2$ stretched in the $y$ direction on its clamped edge by $5\%$ (roughly the order in our MD simulations). The in-plane stress fields $\sigma_{yy}$ and $\sigma_{xx}$ both show a ``crescent moon" domain close to the clamped boundary. But $\sigma_{yy}$ is positive while $\sigma_{xx}$ is negative, which means this area is stretched in the $y$ direction but compressed in the $x$ direction. The high tension in the ``crescent moon" region irons out wrinkles, making it close to horizontal, consistent with the finding that the middle line in Fig.~\ref{fig:profile} starts out flat. The ``crescent moon" is also observed in MD simulations (see Supplementary Movies). Above some threshold then, the compression in the $x$ direction may drive an Euler-type buckling.

We now develop an analytic model that predicts the required conditions for tilt and the associated buckling. We use the thermalized elastic sheet as our reference state and choose the coordinates such that the thermalized sheet occupies the  region $0 \leq x \leq L'$ and $-W'/2 \leq y \leq W'/2$, and is clamped at $x=0$. 
Note that $W'<W_0$ and $L'<L_0$. The deformation from the reference state is described by in plane displacements $u_x(\mathbf{r})$ and $u_y(\mathbf{r})$, and an out-of-plane deflection $h(\mathbf{r})$.
The elastic energy of the system is \cite{Landau}
\begin{equation}\label{energy}
E=\int d^2r \left[\frac{\kappa_R}{2}(\nabla h)^2 + \mu_R u^2_{ij} + 
\frac{1}{2} \lambda_R u^2_{kk}\right],
\end{equation}
where the strain tensor $u_{ij}=\frac{1}{2}(\frac{\partial u_i}{\partial x_j}+\frac{\partial u_j}{\partial x_i} + \frac{\partial h}{\partial x_i}\frac{\partial h}{
\partial x_j})$. 
Thermal fluctuations renormalize the elastic moduli so that they become (strongly) scale-dependent   \cite{Nelson&Peliti, guitter,Aronovitz&Lubensky, LeDoussal&Radzihovsky, Bowick1996, Kosmrlj&Nelson2016}: $\kappa_R(L_0)\sim\kappa \left( \frac{L_0}{l_{th}}\right)^\eta$ and $Y_R(L_0)\sim Y\left( \frac{L_0}{l_{th}}\right)^{-\eta_u}$, where $\eta\approx0.8$ and $\eta_u\approx0.4$.
All lengths are measured in units of the thermal length scale, $l_{th}$, defined as the minimum length scale above which thermal fluctuations significantly renormalize the elastic moduli.  It is given by $l_{th}=\sqrt{\frac{16\pi^3\kappa^2}{3k_B TY}}$, where $Y=4\mu(\mu+\lambda)/(2\mu +\lambda)$ is the bare Young's modulus in the continuum model. Although clamping may suppress thermal fluctuations in a small region near the clamped boundary and give rise to spatial and strain-dependent elastic moduli \cite{LOPEZPOLIN2017}, we assume for simplicity that the renormalized elastic moduli are uniform and strain-independent. Corrections to uniformity will be higher order terms so one can think of this as a Hookean approximation, even though thermalized sheets do display other non-Hookean behavior \cite{Bowick2017, Nicholl2017}.  

\begin{figure}
\centering
\includegraphics[width=8.6cm]{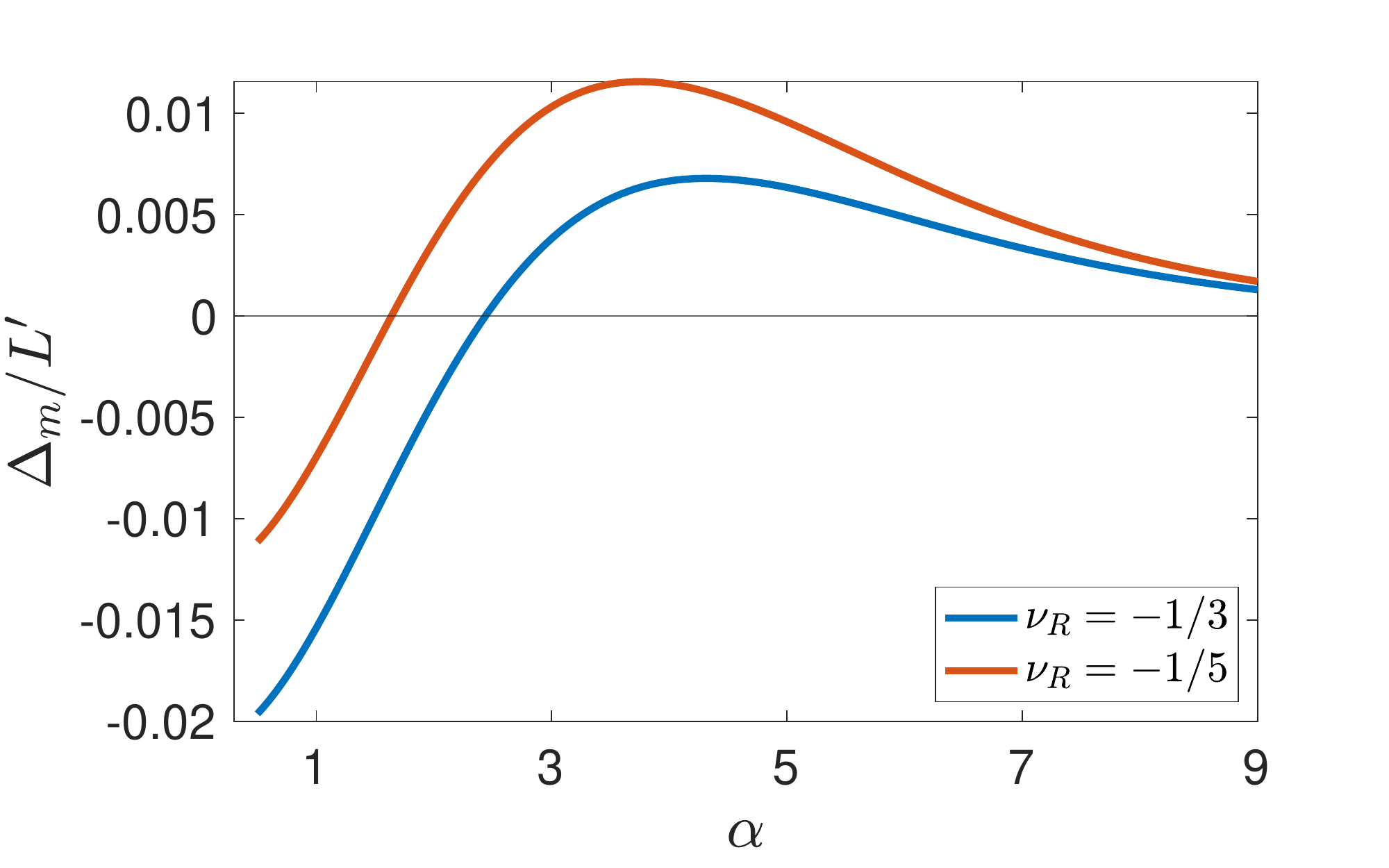}
\caption{(Color online) Compression of middle strip $\Delta_m/L'$ as a function of aspect ratio $\alpha$ using Eq.~(\ref{eq:delta}), with $\epsilon=0.05$ and two different renormalized Poisson ratios.}
\label{fig:displacement}
\end{figure}

Clamping imposes a boundary condition $u_x(0,y)=0$. 
It also fixes the left edge to the zero-temperature width $W_0>W'$, imposing stretching on the reference state:
\begin{equation}\label{stretch condition}
u_y(0,\frac{W'}{2})=-u_y(0,-\frac{W'}{2})=\frac{W_0-W'}{2}\equiv \frac{\epsilon}{2}W_0.
\end{equation}
The extension ratio $\epsilon=(W_0-W')/W_0$ is approximately given by \cite{Kosmrlj&Nelson2016}
\begin{equation}\label{eq:epsilon}
\epsilon \approx \frac{1}{8\pi}\frac{k_B T}{\kappa}\left[\eta^{-1}-\eta^{-1}(l_{th}/L_0)^{\eta}+ \text{ln}(l_{th}/a) \right].
\end{equation}
We find it useful to double our system to the region $-L' \leq x \leq L'$ and $-W'/2 \leq y \leq W'/2$ by reflecting it about the $y$ axis. The originally clamped edge is no longer on the boundary in this doubled system, and $u_x(0,y)=0$ is automatically satisfied by symmetry. We consider a narrow strip with length $2L'$ around $y=0$ and call it the middle strip. To determine the compression $\Delta_m$ of the middle strip we examine the system from its flat ($h=0$) phase in Eq.~(\ref{energy}).
The energy functional gives the equilibrium equation for the in- plane stress $\partial_i \sigma_{ij}=0$ \cite{Landau}. On the left and right edges of the doubled system, we impose the strong traction free boundary condition $\sigma_{xx}(\pm L',y)=0$ and a weak boundary condition $\int\sigma_{xy}(\pm L',y)dy=0$. The latter boundary condition makes the problem analytically solvable. To model the stretching effect from clamping on the reference state we use a delocalized stress on the top and bottom edges in place of a highly localized stress at $x=0$.   Specifically, we impose $\sigma_{xy}(x,\pm W'/2)=0$, and $\sigma_{yy}(x,\pm W'/2)=f\,\text{cos}(\pi x/2L')$, where $f$ is determined self-consistently by condition Eq.~(\ref{stretch condition}) on the displacement. We then compute the stress fields using the Airy stress function method (see Supplementary Material \cite{Supp} for more details). After applying the stress-strain relation, we obtain $u_{xx}(x,0)$, which we integrate to obtain the compression $\Delta_m$ of the middle strip at $y=0$. To first order in small $\epsilon$ we find
\begin{widetext}
\begin{eqnarray}
\Delta_m = -2 u_x(L',0)=\!\frac{L_0\alpha \epsilon}{2\text{sinh}^2(\frac{\pi\alpha}{4})}\!\left[\frac{\pi\alpha}{4}\text{cosh}(\frac{\pi\alpha}{4})(1\!+\!\nu_R)\!
-\!\text{sinh}(\frac{\pi\alpha}{4})(1\!-\!\nu_R)\right]
\label{eq:delta}
\end{eqnarray} 
\end{widetext}
where $\nu_R=\frac{\lambda_R}{2\mu_R +\lambda_R}$ is the renormalized Poisson ratio. We observe that $\Delta_m$ crosses from negative to positive at some threshold aspect ratio (Fig.~\ref{fig:displacement}). This is due to two competing effects. The tensile stress $\sigma_{yy}$ from clamping tends to extend the middle strip because of an overall negative Poisson ratio of the reference state \cite{LeDoussal&Radzihovsky, Bowick1996}. The compressive stress $\sigma_{xx}$, in contrast, tends to compress the middle strip. Our calculation shows that the former dominates for small aspect ratio, extending the middle strip ($\Delta_m<0$), and the latter dominates for a window of higher aspect ratios, allowing for buckling. As expected, a less negative Poisson ratio reduces the effect of $\sigma_{yy}$ and favors compression of the middle strip. For even larger aspect ratio, the two factors balance each other and $\Delta_m$ approaches zero from above, which implies no buckling.

To estimate the critical compression $\Delta_c$ above which the flap buckles we use a one dimensional model for the middle strip. Dropping the $y$ derivatives and $u_y$ in Eq.~(\ref{energy}), we have an energy density functional
\begin{widetext}
\begin{eqnarray}
\mathcal{E}[u_x,h]=\frac{\kappa_R}{2}\!\int\! dx\!\left(\frac{d^2 h}{dx^2}\right)^{\!\!2} \!+\frac{Y_R}{2(1-\nu_R^2)}\!\int\!dx\! \left[\frac{du_x}{dx}+\frac{1}{2}\!\left(\frac{dh}{dx} \right)^{\!2} \right]^{\!2}
\label{E/w}
\end{eqnarray} 
\end{widetext}
with the anti-periodic boundary condition on the displacement $u_x(-L')=-u_x(L')=\Delta/2$.
Integrating out the displacement field gives an effective energy density in terms of $h$ alone. The detailed calculation is given in the Supplemental Material. The result reads
\begin{widetext}
\begin{eqnarray}
\mathcal{E}_{eff}\left[h\right]=&-\text{log}\left[ \int Du_x e^{-\mathcal{E}[u_x,h]}\right]
= \frac{\kappa_R}{2}\!\int\!dx\left( \frac{d^2 h}{dx^2}\right)^{\!2} \!- \frac{Y_R}{2(1-\nu_R^2)}\frac{\Delta}{2L'}\int  \! dx\left(\frac{dh}{dx} \right)^{\!2}
 + \frac{Y_R}{2(1-\nu_R^2)} \frac{1}{8L'}\int\!\!\int\!dxdx' \left(\frac{dh}{dx} \right)^{\!2}\!\left(\frac{dh}{dx'} \right)^{\!2}.
\end{eqnarray} 
\end{widetext}
A generalization of this result, including the quartic term, to a circular plate can be found in \cite{shankar2021}.
Using a mean field variational function (up to an arbitrary constant) $h(x)=H\,\text{cos}\left(\frac{\pi x}{2L'}\right)$, where $H$ serves as the buckling order parameter, yields a critical compression 
\begin{equation}
\Delta_c=\frac{\pi^2}{2L'}\frac{\kappa_R(1-\nu_R^2)}{Y_R}.  
\label{eq:delta_c}
\end{equation}
Setting $\Delta_m=\Delta_c$, a combination of Eq.~(\ref{eq:epsilon}), Eq.~(\ref{eq:delta}) and Eq.~(\ref{eq:delta_c}), gives the phase boundary between the horizontal and tilted phases and is shown with thick white lines in Fig.~\ref{fig:order}. 
The result shows that the tilt phase exists for a finite window of aspect ratios, consistent with our MD simulations. 
The two boundary lines, however, merge at a much lower temperature than the simulation results.
This is an expected overestimation of the tilt phase since our theory only analyzes buckling of the middle strip, which is under the highest compressive stress.
We also note that we have used a constant $\nu_R=-1/3$ which is the universal Poisson ratio for an infinitely sized thermal sheet \cite{LeDoussal&Radzihovsky, Bowick1996}.
Finite-size effects and the suppression of thermal fluctuations from clamping may shift $\nu_R$ and even introduce spatial and strain dependence. 

 The observation that tilt is only present for $kT\geq 0.4 \mbox{eV}$ in MD simulations is a non-universal result of the small system size, and we expect that larger systems favor tilt. Equation (\ref{eq:delta}) suggests that $\Delta_m\sim L_0$, and Eq.~(\ref{eq:delta_c}) suggests that $\Delta_c\sim L_0^{-1+\eta+\eta_u}\approx L_0^{\,\,0.2}$. The amount of compression of the middle strip therefore grows much faster than the critical compression required for tilting as system size increases. In the simulation setup, $L_0=20a$ corresponds to a length of about $50\text{\r{A}}$, but graphene samples in experiments can have lengths over $10\mu m$ \cite{Blees2015}, two thousand times larger than our system. Indeed, we observe tilt with $L_0=60a$ and $\alpha=5$ at $k_BT=0.1eV$, which is an experimentally feasible temperature much lower than the melting temperature of graphene.

In conclusion, we have shown via MD simulations and theory that thin thermalized elastic sheets with one end clamped spontaneously tilt, with respect to the horizontal, for a range of aspect ratios and for sufficiently high temperature. Clamping is shown to induce a  stretching force along the clamped edge which causes a transverse compression that can drive Euler-type buckling. An analytic model, consistent with the simulation results, is developed which predicts the tilt phase diagram.  We hope our work will stimulate experiments which exploit the geometric control of the mechanical behavior of thermalized 2D-metamaterials or other realizations of thermalized elastic sheets. 

\textit{Acknowledgments.} Z.C and D.W contributed equally to this work. The authors thank Rastko Sknepnek for collaboration at the beginning of this work. M.J.B thanks David R. Nelson for many years of discussion on graphene statistical mechanics. Z.C thanks Christopher Jardine and Paul Hanakata for helpful discussions. This research was supported in part by the National Science Foundation under Grant No. NSF PHY-1748958. D.W. acknowledges the support from the National Natural Science Foundation of China (Grant No. 11904265), the Hubei Provincial Natural Science Foundation (Grant No. ZRMS2020001084) and the Fundamental Research Funds for the Central Universities (Grant No. 2042020kf0033). Use was made of computational facilities purchased with funds from the National Science Foundation (CNS-1725797) and administered by the Center for Scientific Computing (CSC). The CSC is supported by the California NanoSystems Institute and the Materials Research Science and Engineering Center (MRSEC; NSF DMR 1720256) at UC Santa Barbara.

\bibliography{refs}
\end{document}